\documentclass[12p]{article}
 \usepackage{dcolumn}

\usepackage{latexsym}
\usepackage{cite}
\usepackage{color}
\usepackage{amsmath}
\usepackage{epsfig}
\usepackage{hyperref}

\newcommand{\ortala}[1]{\begin{center}#1\end{center}}

\newcommand{\sandd}[1]{\left\langle #1\right\rangle}
\newcommand{\sanddr}[1]{\left\langle\left\langle #1\right\rangle\right\rangle_r}

\newcommand{\summ}[3]{{{\underset{#1 }{\overset{#2}{\displaystyle\sum}}}#3}}

\newcommand{\re}[1]{(\ref{#1})}

\newcommand{\eq}[2]{\begin{equation}\label{#1}  #2\end{equation}}

\newcommand{\paran}[1]{\left(#1\right)}

\newcommand{\sch}[1]{Schrodinger}
\newcommand{\komb}[2]{\paran{\begin{array}{c} #1 \\ #2 \end{array}}}

\newcommand{\sanddrtek}[1]{\left\langle\left\langle 
#1\right\rangle\right\rangle_{r}}
\linethickness{0.6pt}

\begin{document}

\ortala{\textbf{Phase Diagrams of Generalized Spin-S Magnetic Binary Alloys}}

\ortala{G\"ul\c{s}en  Karakoyun\footnote{gulsennkarakoyun@gmail.com}}
\ortala{\textit{The Graduate School of Natural and Applied Sciences, Dokuz 
Eyl{\"u}l University, Tr-35160 {\.I}zmir, Turkey}}

\ortala{\"Umit Ak\i nc\i\footnote{umit.akinci@deu.edu.tr}}
\ortala{\textit{Department of Physics, Dokuz Eyl\"ul University,
TR-35160 Izmir, Turkey}}

\section{Abstract}\label{Abstract}

Critical properties of the generalized spin-S 
magnetic binary alloys represented by $A_{c}B_{1-c}$ have been investigated within the framework of 
EFT. By inspecting the evolution of the phase diagrams with the concentration for several spin values, general results have been obtained. Obtained results cover the results obtained for special cases in the literature. Type of the transition (first/second order), as well as the presence of the tricritical point have been determined for general spin models. It has also been shown that, the same critical concentration value exist in the system, regardless of the spin value for binary alloy consist of half integer-integer spin alloy.

\textbf{Keywords:} Bimodal random field, binary alloy, effective field theory

\section{Introduction}\label{introduction}

Basic spin models, such as the Ising model, have an important role as a pioneering 
model which applies to some 
complex and realistic magnetic systems. The spin-$1$ Blume Capel (BC) model \cite{ref1,ref2}, which has
the single ion anisotropy, draws great deal of attention with its multicritical phenomena. Demand for expanded
Ising models with higher spin values has increased day by day, due to the requirement to examine the magnetic 
properties of real physical systems. In this context, generalized spin-$S$ models have been investigated by means
of several methods  such as mean field theory (MFT) \cite{ref3} for a spin glass \cite{ref4}; effective field
theory (EFT) \cite{ref5,ref6,ref7,ref8,ref_9_1} for site diluted \cite{ref9}, transverse \cite{ref10} and random field 
\cite{ref11} Ising models. Besides,  Monte Carlo (MC) method \cite{ref12,ref13}, series expansion  methods \cite{ref14,ref15}
and pair approximation methods \cite{ref16,ref17} are widely used in the literature.

There are comprehensive works about mixed-spin Ising models which consist of two different spin values in the 
literature. Mixed Ising models with half-integer half-integer  spins have been examined within the framework of EFT \cite{ref18,ref19,ref20,ref21}, MC \cite{ref22,ref23,ref24}, 
exact recursion relations \cite{ref25} and Oguchi approximation \cite{ref26}. No matter which method or lattice model 
has been used, the system holds minimum half integer $s=\pm 1/2$ ordered states at values of  large negative single ion anisotropy parameter 
\cite{ref21,ref22,ref23,ref25,ref26}. When the values of the mixed spin Ising model are selected as integer the  first
order phase transitions and tricritical points have been obtained by using of MFT \cite{ref27}. At the same time, 
the reentrant phenomena has also  been confirmed by the MC method in this study. Besides, integer - half integer  mixed spin Ising model with two different single ion anisotropies have been investigated
by means of MFT \cite{ref28,ref29} and EFT \cite{ref30}. On account of two different single ion anisotropies acting on the 
system, different types of phase diagram were achieved, one pf which contains tricritical points and the first 
order phase transitions whereas  another don't. If we consider the model with the integer-half integer spins 
without anisotropy, study on random magnetic binary
alloy $A_xB_{1-x}$ by MC, it has been concluded that this model is in the same universality class of the two-dimensional pure Ising
model \cite{ref31}. Furthermore a lot of mixed spin systems are modeled where at least one of
the components is generalized spin-$S$. For instance, mixed spin-$1/2$ and spin-$S$ system has been examined by EFT 
\cite{ref32}, high-temperature series expansion \cite{ref33}, free Fermion approximation \cite{ref34} and exact 
calculations \cite{ref35,ref36,ref37}. Notice that, especially these exact results show that integer or half integer 
decorating spin models exhibit qualitatively analogous to critical behavior of standard Ising model
for all planar lattices. Besides, 
generalized spin-$S$ and spin-$S'$ model have been investigated by high-temperature series expansions \cite{ref38}.

Since the phase diagrams of magnetic binary alloys show remarkable interesting properties, many systems have been
studied with different approaches with increasing interest over many years. Ferromagnetic binary magnetic alloy  system of type-A and type-B atoms have been investigated by 
EFT \cite{ref39,ref40}, MC \cite{ref41}, MFT \cite{ref42,ref43}. Perturbation theory was also used for 
extended spin S model \cite{ref44}. On the other hand, the system consisting of different spin values such as 
spin-$1/2$-$1$ has been examined by several methods such as EFT for amophous ferrimagnetic \cite{ref45,ref46,ref47} 
and ferromagnetic \cite{ref48,ref49,ref50,ref51_2} binary alloys; MFT \cite{ref51,ref52,ref53} and MC \cite{ref54} methods. 
Spin-$1/2$-$3/2$ model has been inspected by use of MFT \cite{ref52,ref55} and also within EFT \cite{ref56,ref57}. Phase diagrams of bilayer \cite{ref58} and multilayer \cite{ref59} systems consisting
of spin-$1/2$ and spin-$3/2$ Ising layers  have been studied. The generalized one component form of disordered
ferrimagnetic binary alloys have been obtained on the basis of MFT for $S_A=1/2$ and $S_B>1$ \cite{ref60} and within 
the two frameworks EFT and MFT for $S_A=1/2$ and $S_B>S_A$ \cite{ref56}. Generalization of both spin variables
of binary ferromagnetic alloy has been examined with easy axis and easy plane competing anisotropies by means of molecular
field approximation \cite{ref61}. Phase diagrams of the $AB_pC_{1-p}$ ternary alloy consisting of $S_A=3/2$, $S_B=2$
and $S_C=5/2$ and $S_A=3/2$, $S_B=1$ and $S_C=5/2$ have been established by MFT \cite{ref62}, MC simulations 
\cite{ref63}, respectively. The effect of single-ion anisotropy and concentration on the phase diagrams are
obtained and appearance of multicritical points have been demonstrated. Besides, $S_A=1$, $S_B=3/2$ and $S_C=1/2$ ternary alloys
have been investigated within the two frameworks EFT and MC \cite{ref64}.

It is convenient to use well-known Ising like models, in order to determine the magnetic properties of the disordered
binary alloys. $Fe_{1-q}Al_q$ alloys have been constructed on a basis of a site-diluted Ising spin model by means
of EFT \cite{ref65,ref66}. Mixed-bond (can be considered as random-bond) spin-$1/2$ Ising model for 
$Fe-Mn$ \cite{ref67} and  $Fe-Ni-Mn$ alloys \cite{ref68} have been studied within the EFT. Random-bond Blume
Capel model has been constructed for ternary $(Fe_{0.65}Ni_{0.35})_{1-x}Mn_x$ and $Fe_pAl_qMn_x$ alloys \cite{ref69}. On the other hand, phase diagrams of
transition metal alloys have been extensively examined by both experimental and theoretical points of view based on
the phenomenological models of statistical physics \cite{ref70}.
Besides, experimental implementation of magnetic alloy systems was supported by experimental studies such as
$FeAl_{1-x}Mn_x$ \cite{ref71}, $Co(S_xSe_{1-x})_2$ \cite{ref72}, $Fe_{100-x}B_x$ \cite{ref73}, $Tb_xY_{1-x}$
and $Tb_xGd_{1-x}$ alloys \cite{ref74}.

It can be seen from this short literature, binary magnetic models are still up to date. But general results are still needed.  Most of the studies were done on specific values of the spins. Thus, 
the main aim of this work to determine the multi-critical behavior of the binary alloys by taking both generalized spin-$S$
of type-A and type-B atoms consisting of different spin values. The outline of this paper is as follows: In Section 2
the formulation of generalized spin-$S$ magnetic binary alloy has been constructed by EFT. 
In Section 3, results and discussions are presented and finally Section 4 contains our conclusions.

\section{Model and Formulation}\label{model}

The chemical formula of the binary alloy can be given by $A_cB_{1-c}$. Randomly distributed 
two types of atoms denoted by $A$ and $B$ brought
together with the concentrations  $c$ and $1-c$, respectively. Our investigation will be focused on a general system
which is constituted by  type-A  atoms that have spin-$S_A$ and type-B atoms that have spin-$S_B$ within the Ising model. 

The Hamiltonian of the binary Ising model is given by
\eq{denk1}{\mathcal{H}=-J\summ{<i,j>}{}{\paran{\xi_i \xi_j  \sigma_i \sigma_j+
\xi_i \delta_j \sigma_i s_j+
\delta_i \xi_j s_i \sigma_j+
 \delta_i \delta_j s_i s_j}}-D\summ{i}{}{\paran{\xi_i \sigma_i^2+\delta_i 
s_i^2}},}
where $\sigma_i,s_i$ are the $z$ components of the spin-$S_A$ and spin-$S_B$ 
operators 
and they take the values
$\sigma_i=-S_A,-S_A+1,\ldots, S_A-1,S_A$ and $s_i=-S_B,-S_B+1,\ldots, S_B-1,S_B$, respectively.  
$J>0$ is the   ferromagnetic
exchange interaction between the nearest neighbor spin pairs, $D$ is
the crystal field (single ion anisotropy).
One site of the lattice is occupied by type-A atoms  if $\xi_i=1$ and 
type-B atoms  if $\delta_i=1$. Since there is no vacancy on the lattice, the site occupation number related to the 
site holds the relation $\xi_i+\delta_i=1$. 
The first summation in Eq.
\re{denk1} is over the nearest-neighbor pairs of spins and the second
summation is over all the lattice sites.

Let us concentrate on a site labeled by $0$. All interactions of this site can be represented by $ 
\mathcal{H}_0^{A} / \mathcal{H}_0^{B}$ if the site $0$ is occupied by $A/B$ type atoms, respectively. These terms
can be regarded as local fields acting on a site $0$ and they are given by,

\eq{denk2}{
\mathcal{H}_0^{A}=-\xi_0\sigma_0\left[J\summ{j=1}{z}{\paran{\xi_j \sigma_j + 
\delta_j  s_j }}\right]-\xi_0\paran{\sigma_0}^2 D=-\xi_0\sigma_0\left[E_0^{A}\right]-\xi_0\paran{\sigma_0}^2 D,} 

\eq{denk3}{
\mathcal{H}_0^{B}=-\delta_0 s_0\left[J\summ{\delta=1}{z}{\paran{\xi_j  \sigma_j 
+ \delta_j s_j }}\right]-\delta_0\paran{s_0}^2 D =
-\delta_0 s_0\left[E_0^{B}\right]-\delta_0\paran{s_0}^2 D.} 

We can  use of the exact identities \cite{ref15a} which are given by

$$
m_A=\frac{\sanddr{\xi_0\sigma_0}}{\sandd{\xi_0}_r}=\frac{1}{\sandd{\xi_0}_r}
\sanddr{\frac{Tr_0\xi_0 \sigma_0 \exp{\paran{-\beta
\mathcal{H}_0^{A}}}}{Tr_0\exp{\paran{-\beta \mathcal{H}_0^{A}}}}},
$$

$$
q_A=\frac{\sanddr{\xi_0\sigma_0^2}}{\sandd{\xi_0}_r}=\frac{1}{\sandd{\xi_0}_r}
\sanddr{\frac{Tr_0\xi_0 \sigma_0^2 \exp{\paran{-\beta
\mathcal{H}_0^{A}}}}{Tr_0\exp{\paran{-\beta \mathcal{H}_0^{A}}}}},
$$

\eq{denk4}{m_B=\frac{\sanddr{\delta_0s_0}}{\sandd{\delta_0}_r}=\frac{1}{\sandd{
\delta_0}_r}\sanddr{\frac{Tr_0 \delta_0 s_0 \exp{\paran{-\beta
\mathcal{H}_0^{B}}}}{Tr_0\exp{\paran{-\beta \mathcal{H}_0^{B}}}}},}

$$
q_B=\frac{\sanddr{\delta_0s_0^2}}{\sandd{\delta_0}_r}=\frac{1}{\sandd{\delta_0}
_r}\sanddr{\frac{Tr_0 \delta_0 s_0^2
\exp{\paran{-\beta \mathcal{H}_0^{B}}}}{Tr_0\exp{\paran{-\beta
\mathcal{H}_0^{B}}}}},
$$
where $Tr_0$ is the partial trace over the site
$0$, $\beta=1/\paran{k_B T}$, $k_B$ is Boltzmann constant
and $T$ is the temperature. In order to get the magnetizations ($m_A,m_B$) and the quadrupolar moments 
($q_A,q_B$) of the system, thermal averages (inner 
bracket) and random configurational averages
(bracket with subscript $r$) have to be taken.


From now on, we derive equation related to $m_A$ from Eq. \re{denk4}. 
Derivation of the equations for $q_A,m_B$ and $q_B$ can be 
realizable in the same way. 

By writing Eqs. \re{denk2} in Eq. \re{denk4} and 
performing  partial trace operations by using identity $
e^{\xi x}=\xi e^{x}+1-\xi,
$ (where $x$ is any real number and $\xi=0,1$)  we can obtain an expression  in a closed form as

\eq{denk5}{
\frac{\sanddr{\xi_0 \sigma_0}}{\sandd{\xi_0}_r}=
\sanddr{f_m^A\paran{ E_0^{A}}},} where the function is given by \cite{ref16a}

\eq{denk6}{f_m^A\paran{x,D}=\frac{\summ{k=-S_A}{S_A}{}k\exp{\paran{\beta D 
k^2}\sinh{\left(\beta k x\right)}}}{\summ{k=-S_A}{S_A}{}\exp{\paran{\beta D 
k^2}\cosh{\left(\beta k x\right)}}},
}

By using differential operator technique \cite{ref17a}, Eq. \re{denk5} can be 
written as

\eq{denk7}{
\frac{\sanddr{\xi_0 \sigma_0}}{\sandd{\xi_0}_r}=\sanddrtek{e^{E_0^{A}\nabla}}f_m^A(x)|_{x=0},
} where  $\nabla$ represents the differential with respect to $x$. 
The
effect of the differential operator $\nabla$ on an arbitrary function $F$ is
given by
\eq{denk8}{\exp{\paran{a\nabla}}F\paran{x}=F\paran{x+a},} with arbitrary 
constant $a$.

By writing $E_0^A$ from Eq. \re{denk2} in Eq. \re{denk7} we can expand 
the exponential differential operator by using the approximated Van der 
Waerden identity \cite{ref18a}, which is given by

\eq{denk9}{
\exp{\paran{aS}}=\cosh{\paran{a\eta}}+\frac{S}{\eta} \sinh{\paran{a\eta}},
} where $\eta^2=\sandd{S^2}$ and $S$ is the spin eigenvalue.

By using Eq. \re{denk9} in Eq. \re{denk7} with the identity $e^{\xi x}=\xi e^{x}+1-\xi,
$ 
we can obtain magnetization ($m_A$) as

\eq{denk10}{
m_A=\summ{p=0}{z}{}\summ{q=0}{z-p}{}\summ{r=0}{p}{}\summ{s=0}{z-q-r}{}\summ{t=0}{q+r}{}
C_{pqrst}(-1)^tc^{z-p}\paran{1-c}^p \paran{\frac{m_A}{\eta_A}}^q\paran{\frac{m_B}{\eta_B}}^r 
f_m^A\paran{\left[z-2s-2t\right] J,D} 
}
where $\eta_A^2=q_A=\sandd{\sigma^2}$, and 
\eq{denk11}{
C_{pqrst}=\frac{1}{2^z}\komb{z}{p}\komb{z-p}{q}\komb{p}{r}\komb{z-q-r}{s}\komb{q+r}{t}.
}

By using the same procedure between Eqs. \re{denk5} and \re{denk10} we can obtain other quantities as,

\eq{denk12}{
q_A=\summ{p=0}{z}{}\summ{q=0}{z-p}{}\summ{r=0}{p}{}\summ{s=0}{z-q-r}{}\summ{t=0}{q+r}{}
C_{pqrst}(-1)^tc^{z-p}\paran{1-c}^p \paran{\frac{m_A}{\eta_A}}^q\paran{\frac{m_B}{\eta_B}}^r 
f_q^A\paran{\left[z-2s-2t\right] J,D} 
} 

\eq{denk13}{
m_B=\summ{p=0}{z}{}\summ{q=0}{z-p}{}\summ{r=0}{p}{}\summ{s=0}{z-q-r}{}\summ{t=0}{q+r}{}
C_{pqrst}(-1)^tc^{z-p}\paran{1-c}^p \paran{\frac{m_A}{\eta_A}}^q\paran{\frac{m_B}{\eta_B}}^r 
f_m^B\paran{\left[z-2s-2t\right] J,D} 
}

\eq{denk14}{
q_B=\summ{p=0}{z}{}\summ{q=0}{z-p}{}\summ{r=0}{p}{}\summ{s=0}{z-q-r}{}\summ{t=0}{q+r}{}
C_{pqrst}(-1)^tc^{z-p}\paran{1-c}^p \paran{\frac{m_A}{\eta_A}}^q\paran{\frac{m_B}{\eta_B}}^r 
f_q^B\paran{\left[z-2s-2t\right] J,D} 
} Here the functions are defined as \cite{ref16a},

\eq{denk15}{f_q^A\paran{x,D}=\frac{\summ{k=-S_A}{S_A}{}k^2\exp{\paran{\beta D 
k^2}\cosh{\left(\beta k x\right)}}}{\summ{k=-S_A}{S_A}{}\exp{\paran{\beta D 
k^2}\cosh{\left(\beta k x\right)}}}.
}

\eq{denk16}{f_m^B\paran{x,D}=\frac{\summ{k=-S_B}{S_B}{}k\exp{\paran{\beta D 
k^2}\sinh{\left(\beta k x\right)}}}{\summ{k=-S_B}{S_B}{}\exp{\paran{\beta D 
k^2}\cosh{\left(\beta k x\right)}}},
}

\eq{denk17}{f_q^B\paran{x,D}=\frac{\summ{k=-S_B}{S_B}{}k^2\exp{\paran{\beta D 
k^2}\cosh{\left(\beta k x\right)}}}{\summ{k=-S_B}{S_B}{}\exp{\paran{\beta D 
k^2}\cosh{\left(\beta k x\right)}}}.
}

By solving the system of nonlinear equations system  given by  Eqs.
\re{denk10} and \re{denk12}-\re{denk14}  by using the coefficients given in 
\re{denk11} and the definitions of functions given in Eqs. \re{denk6} and \re{denk15}-\re{denk17}, we
can obtain the total magnetization ($m$) and quadrupolar moment ($q$) of the system via
\eq{denk18}{
m=cm_A+\paran{1-c}m_B, \quad q=cq_A+\paran{1-c}q_B.
}

The second order 
critical temperature of the system can be obtained by solving  
linearized versions of the equation system in $m_A$ and $m_B$.

\section{Results and Discussion}\label{results}

All results have been obtained for $z=3$ honeycomb lattice throughout  this work. We 
will utilize scaled (dimensionless) quantities as,

\eq{denk18}{ d=\frac{D}{J},t=\frac{k_BT}{J}. }

We investigate the effect of crystal field and concentration on the critical behavior of the general 
spin-S binary alloy model in four distinct parts:   half 
integer-half integer model, integer - integer model, integer - half 
integer spin model and half integer - integer spin model.
In 
this study, we use $S_{A}<S_{B}$.

\subsection{Half integer - half integer spin model}

\begin{figure}[!h]
\epsfig{file=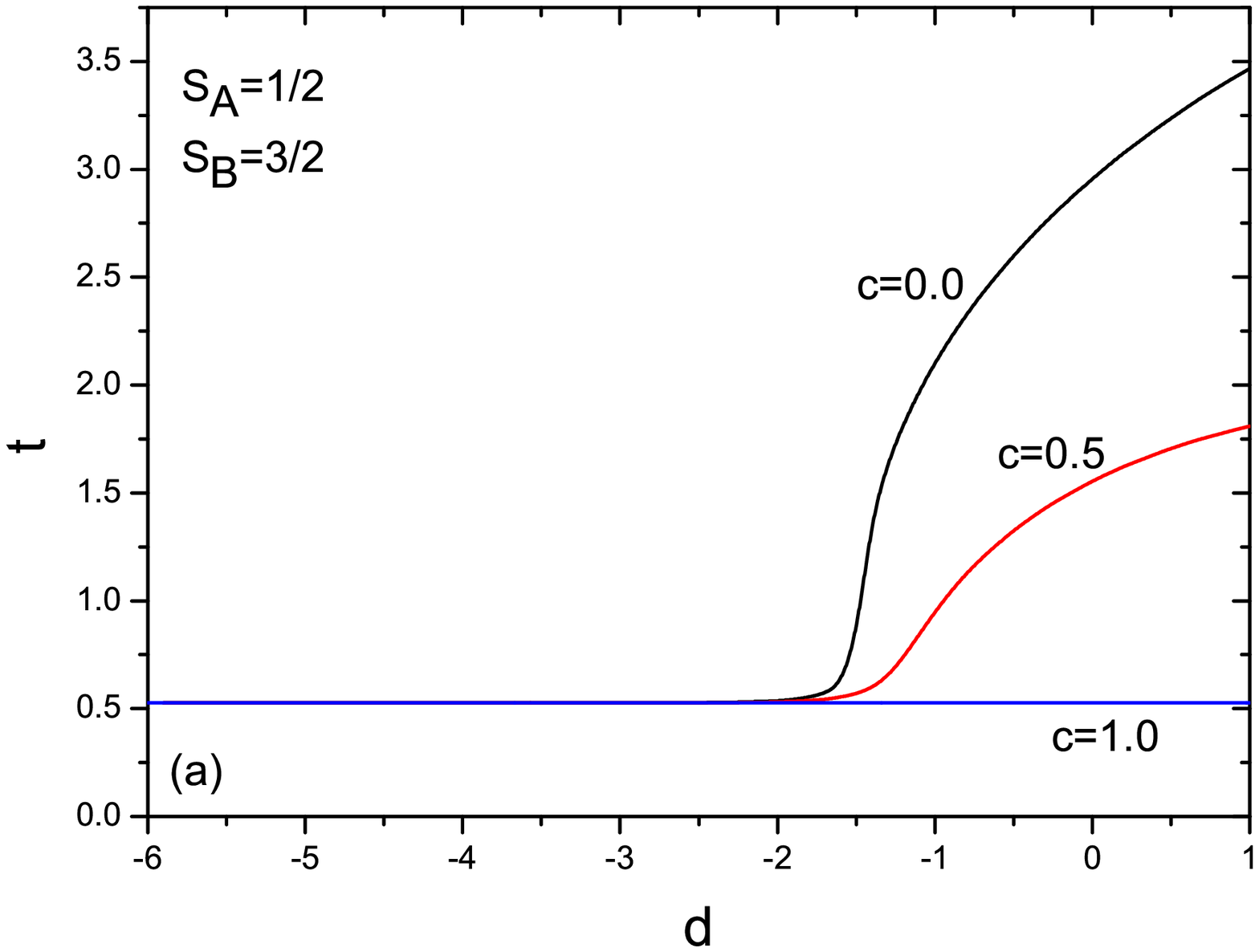, width=8cm}
\epsfig{file=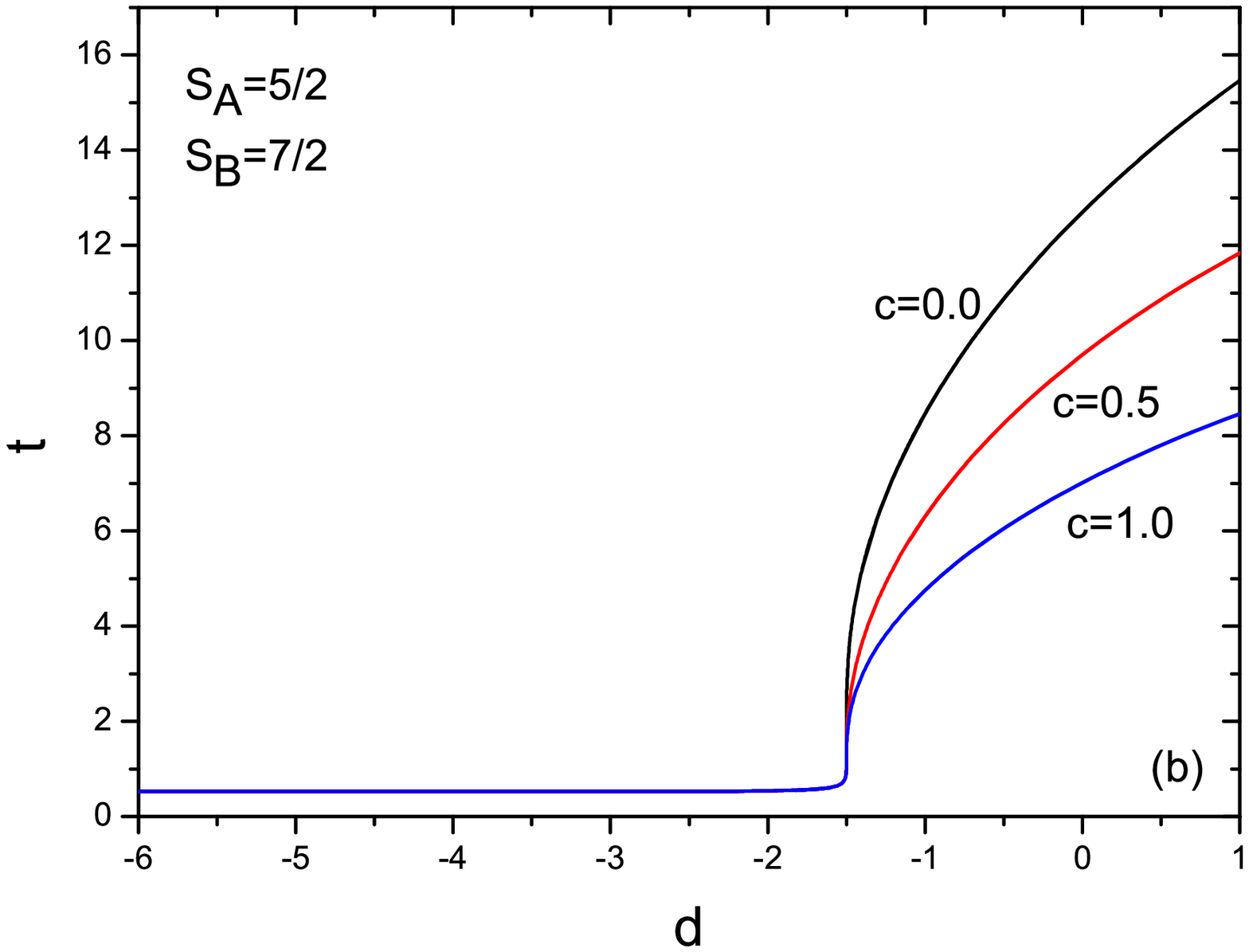, width=8cm}
\caption{Variation of the critical temperature with crystal field parameter for 
selected 
values of the concentration $c=0.0$, $c=0.5$, $c=1.0$. Spin values of type-A 
and type-B atoms chosen as (a) $S_A=1/2$, $S_B=3/2$ and (b) $S_A=5/2$, 
$S_B=7/2$.
Solid lines represent to the second order transitions.}\label{sek1}
\end{figure}

First we choose both spin-A and spin-B values as half integer - half integer.
The phase diagrams in $(d,t)$ plane of the system can be seen in 
Fig. \ref{sek1} with selected values of the concentrations $c=0.0$, $c=0.5$ and $c=1.0$. 
We specify spin values as $S_{A}=1/2$, $S_{B}=3/2$ in Fig \ref{sek1} (a) and $S_{A}=5/2$, 
$S_{B}=7/2$ in Fig \ref{sek1} (b). Note that, in the limiting cases $c = 0$ and $c = 1$, 
all of the magnetic atoms of the system consists of spin-$S_B$ and 
spin-$S_A$, respectively. Phase diagram of the binary alloy system 
evolves according to these limiting cases, when the concentration increases. As we can see 
in Fig \ref{sek1} (a), the
critical 
temperature of the spin-$3/2$ model is higher than the critical temperature of 
the spin-$1/2$ model, for the positive values of crystal field parameter. When the
concentration value increases from $0$ to $1$, critical temperatures decrease 
to the critical temperatures of the spin-$1/2$ model. The system exhibits only second order phase transitions. 
When $d$ takes large negative values, critical temperatures 
gradually merge and phase diagram of the system evolves to a parallel line with respect to 
the crystal field axis. Due to the effect of the large negative crystal field 
values, spins align the minimum value of their possible eigenvalues. In other words, the system prefer $\pm 1/2$ 
states in order to provide minimization of the free energy of the system. 
Ferromagnetic phase holds for the region below this critical temperature. In Fig \ref{sek1} (b) 
spin values of the system are set as $S_{A}=5/2$ $S_{B}=7/2$. It is clear from this 
figure that, the system has higher critical temperature in comparison to the system consisting of 
$S_{A}=1/2$, $S_{B}=3/2$. While the crystal field 
parameter takes large negative values, critical temperature values converge at 
one point, and concentration is no longer important for certain temperature 
values. We can say that, the behaviors of the critical temperatures obtained in this case are 
compatible with the mixed \cite{ref21,ref25} and
binary alloy systems \cite{ref49,ref50,ref51_2}.

\subsection{Integer - Integer spin model}

\begin{figure}[!h]
\epsfig{file=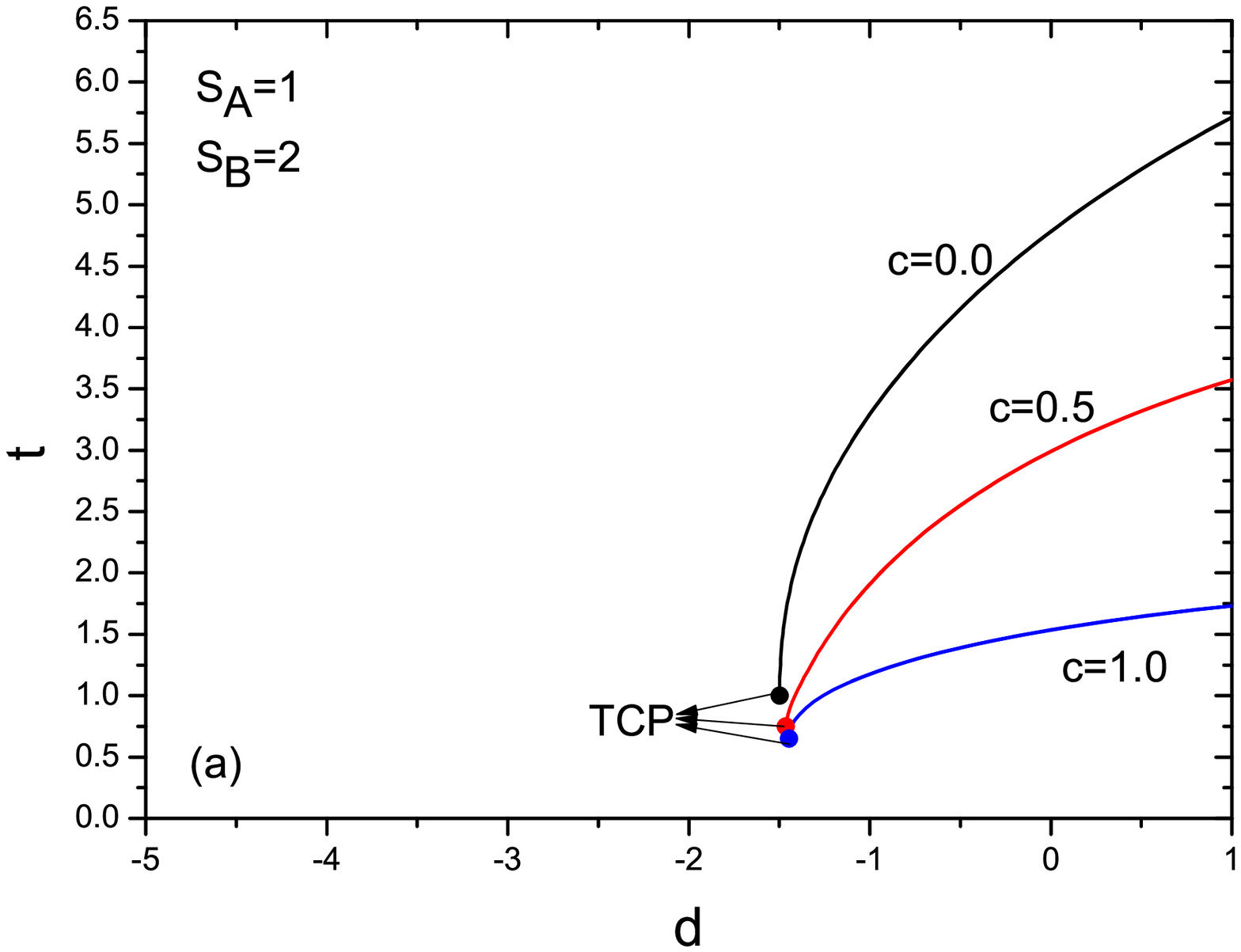, width=8cm}
\epsfig{file=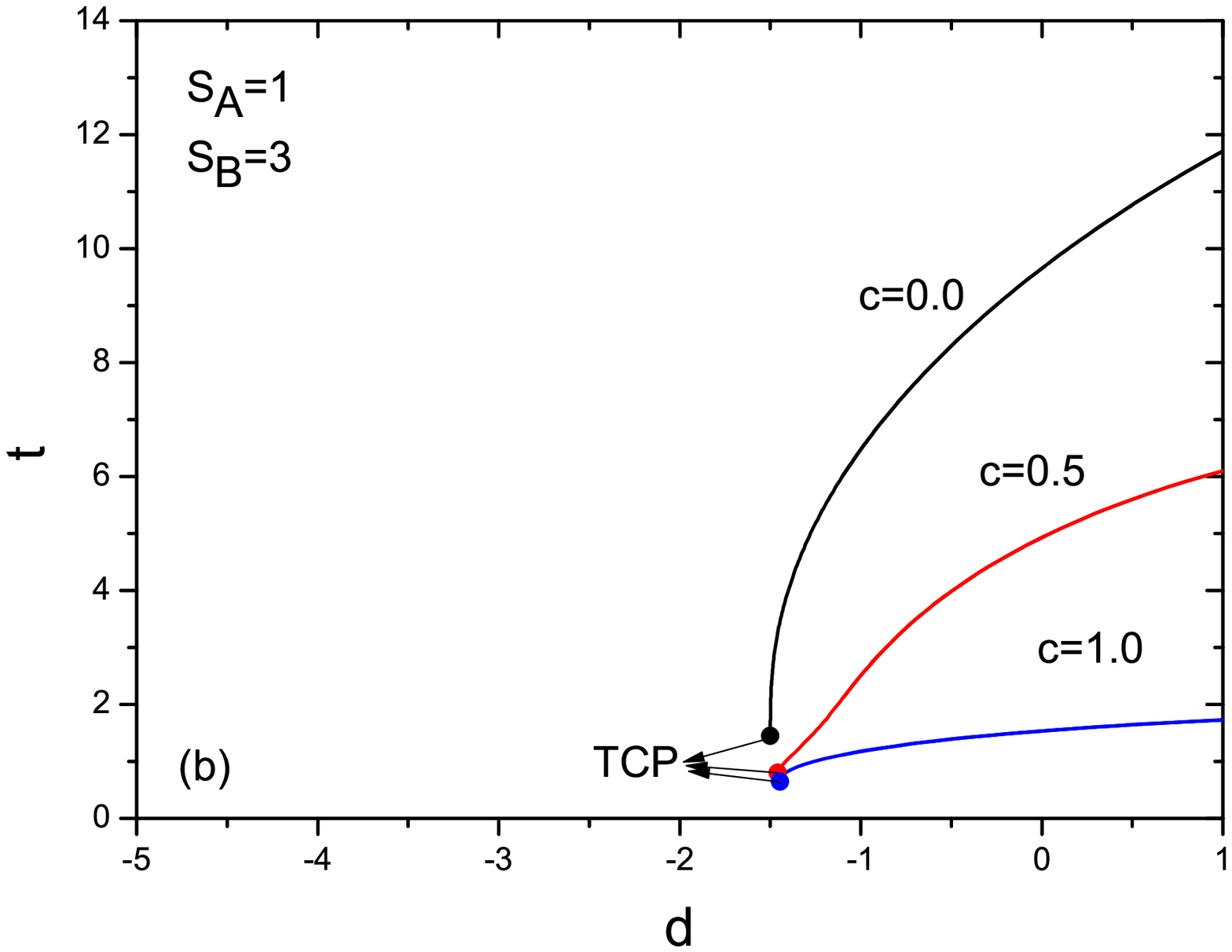, width=8cm}
\caption{Variation of the critical temperature with crystal field parameter for 
selected 
values of the concentration $c=0.0$, $c=0.5$, $c=1.0$. Spin values of type-A 
and type-B atoms are chosen as (a) $S_A=1$, $S_B=2$ and (b) $S_A=1$, $S_B=3$.
Solid lines represent the second order transitions. In figure, TCP stands for 
the tricritical point.}\label{sek2}
\end{figure}

The second case of binary alloy system consists of both spin-A and spin-B 
with integer-integer values. In Fig. \ref{sek2}, the phase diagram in 
$(d,t)$ plane of the system is given for selected values of the concentrations 
$c=0.0$, $c=0.5$ and $c=1.0$. Spin values of the system have been chosen as $S_{A}=1$ 
$S_{B}=2$ in Fig  \ref{sek2} (a). We have examined only second order phase transition 
lines of phase diagram. The system 
exhibits tricritical point (TCP) as it passes from the second order phase 
transition to the first order phase transition. Behavior of the transition lines and 
reentrant phenomena 
in the presence of single-ion anisotropy is in agreement with the mixed spin-$1$ and spin-$2$ Ising system \cite{ref27}. 
As seen in Figs. \ref{sek2}(a) and (b), TCP decreases as the concentration of type-A 
atoms increases. When the crystal field parameter takes large negative values, 
disordered phase appears for ground state, which  consist of  mostly occupied $s=0$ 
states, for all concentrations. If we fixed the spin value of A atoms as $S_{A}=1$ 
and then increase spin value of B atoms such as $S_{B}=3$, then the critical 
temperatures also increases as the concentration of B atoms increases for the positive 
crystal field parameter values ( compare Figs. \ref{sek2} (a) and (b)). Besides it can be seen from Fig. \ref{sek2} (b) that, 
TCP increases as the concentration goes towards zero, as in the case of $S_{A}=1, S_{B}=2$.

\subsection{Integer - Half integer spin model}

\begin{figure}[!h]
\epsfig{file=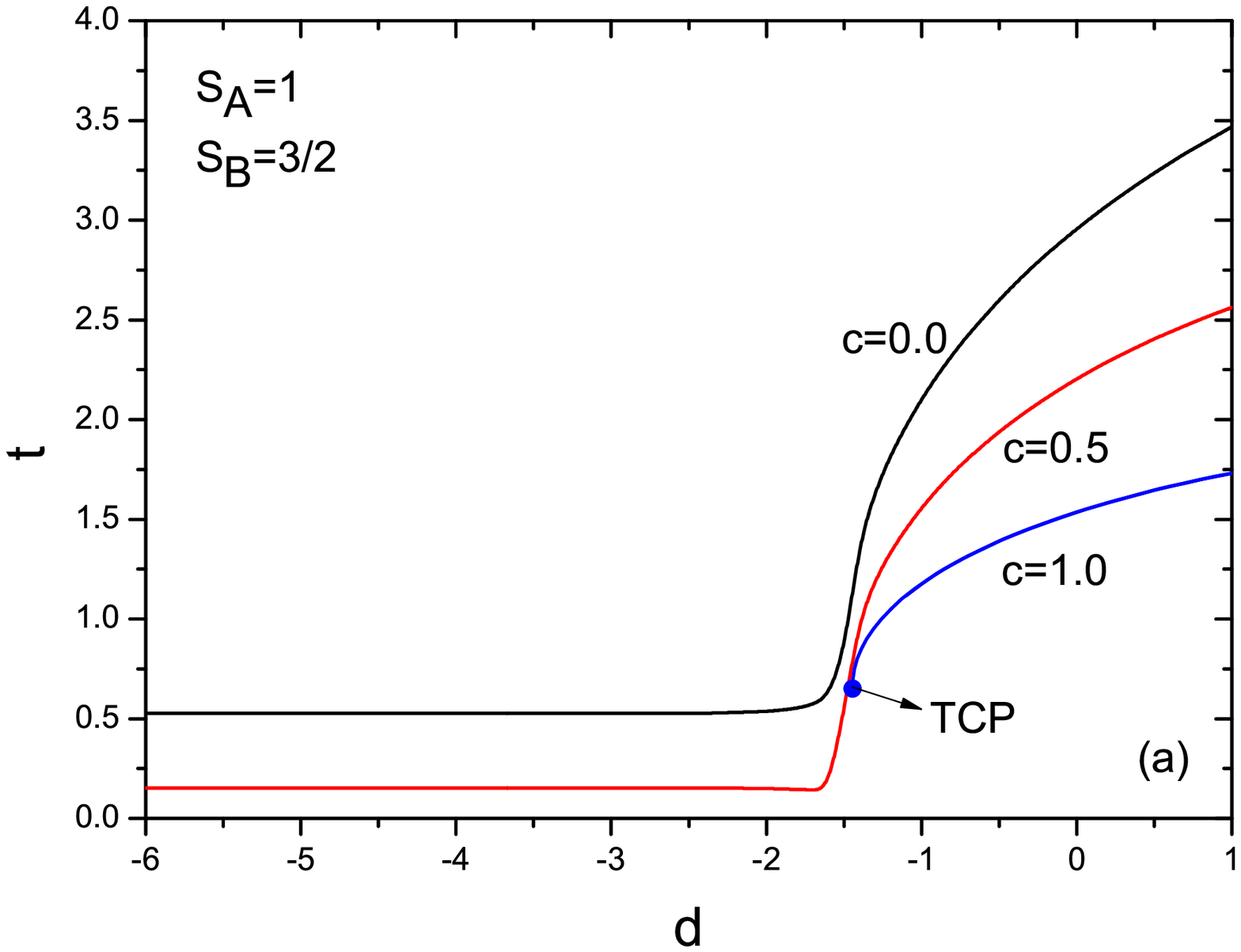, width=8cm}
\epsfig{file=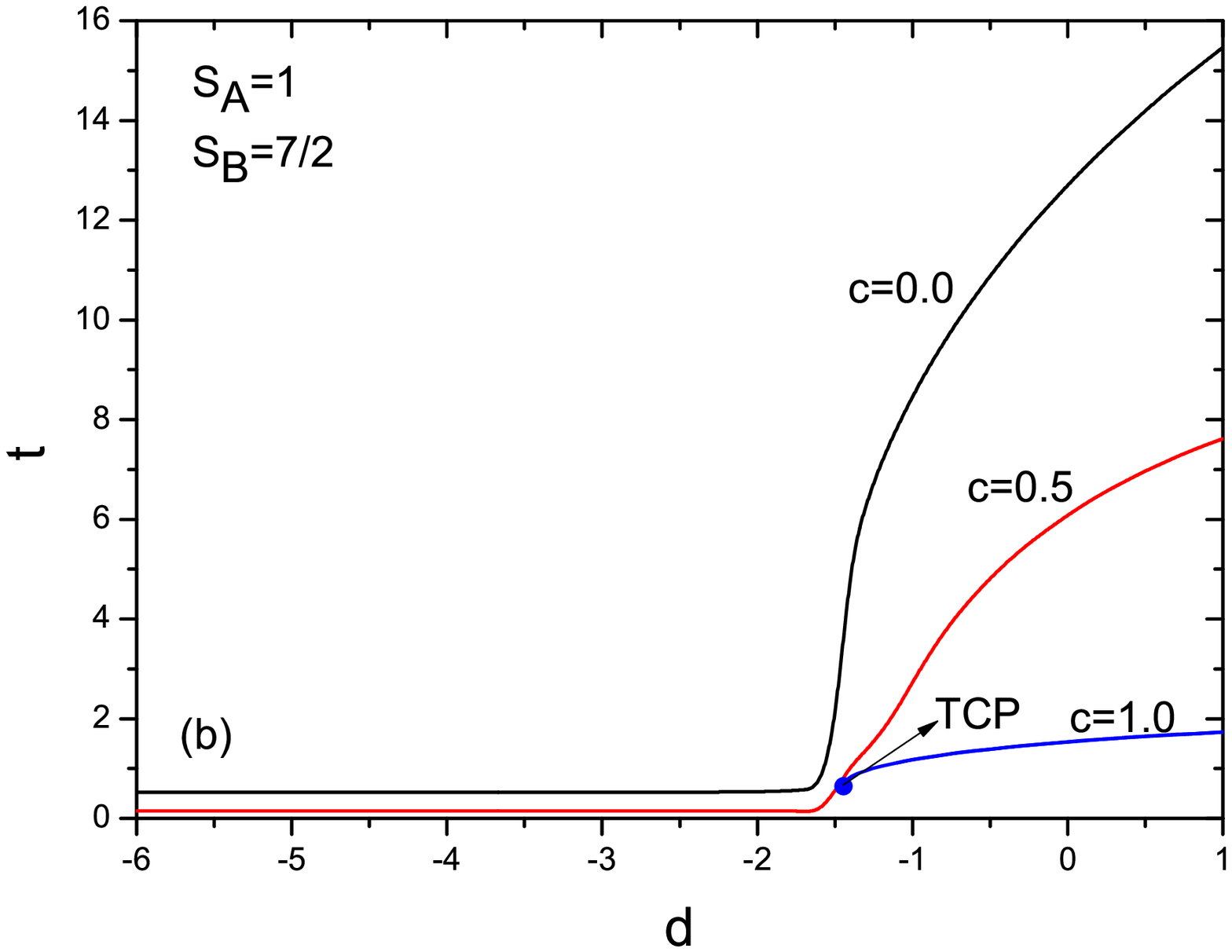, width=8cm}

\centering
\epsfig{file=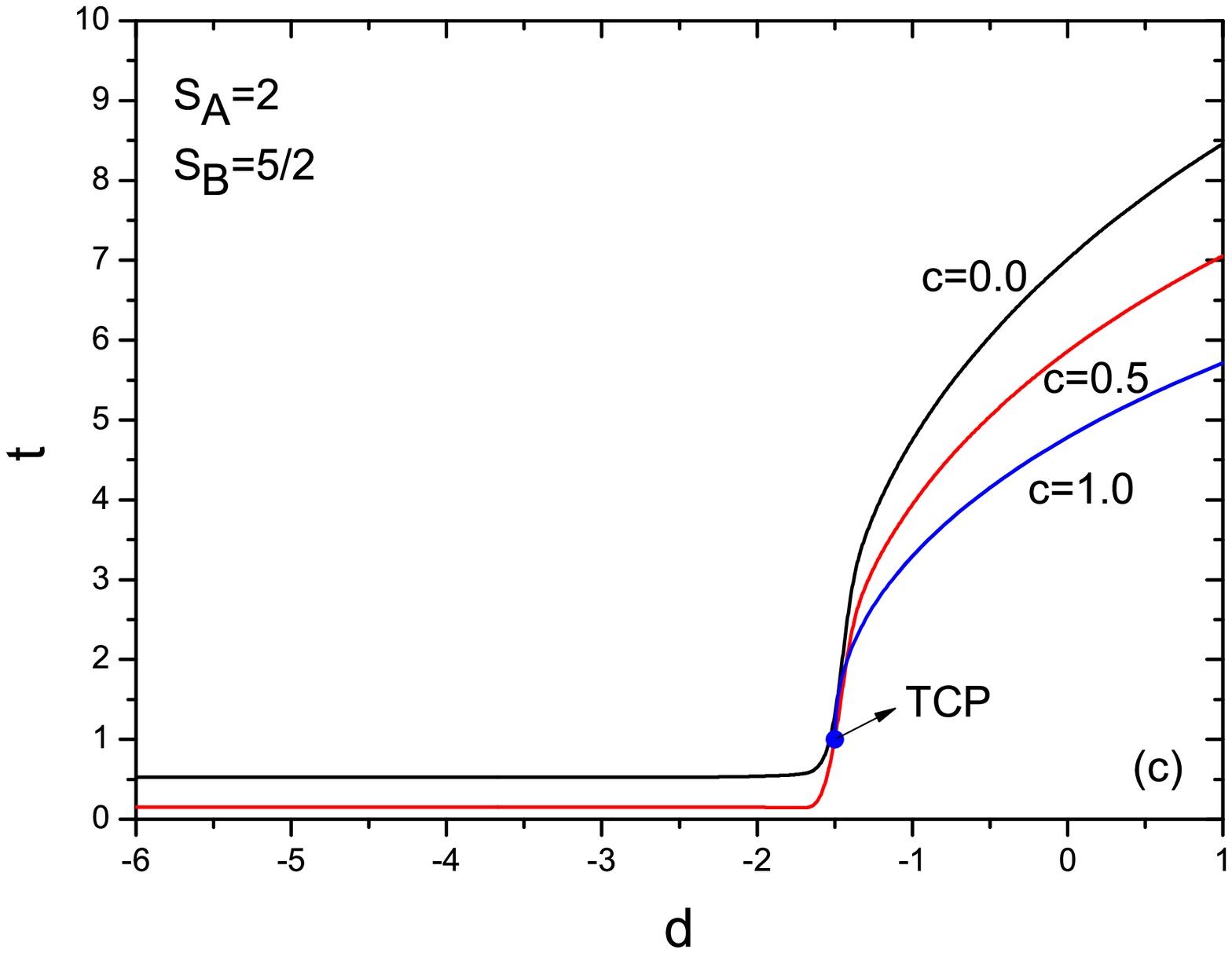, width=8cm}
\caption{Variation of the critical temperature with crystal field parameter for 
selected 
values of the concentration $c=0.0$, $c=0.5$, $c=1.0$. Spin values of type-A 
and type-B atoms are chosen as (a) $S_A=1$, $S_B=3/2$, (b) $S_A=1$, $S_B=7/2$ and 
(c) $S_A=2$, $S_B=5/2$.
Solid lines represent the second order transitions. In figure, TCP stands for 
the tricritical point.}\label{sek3}
\end{figure}

\begin{figure}[!h]
\epsfig{file=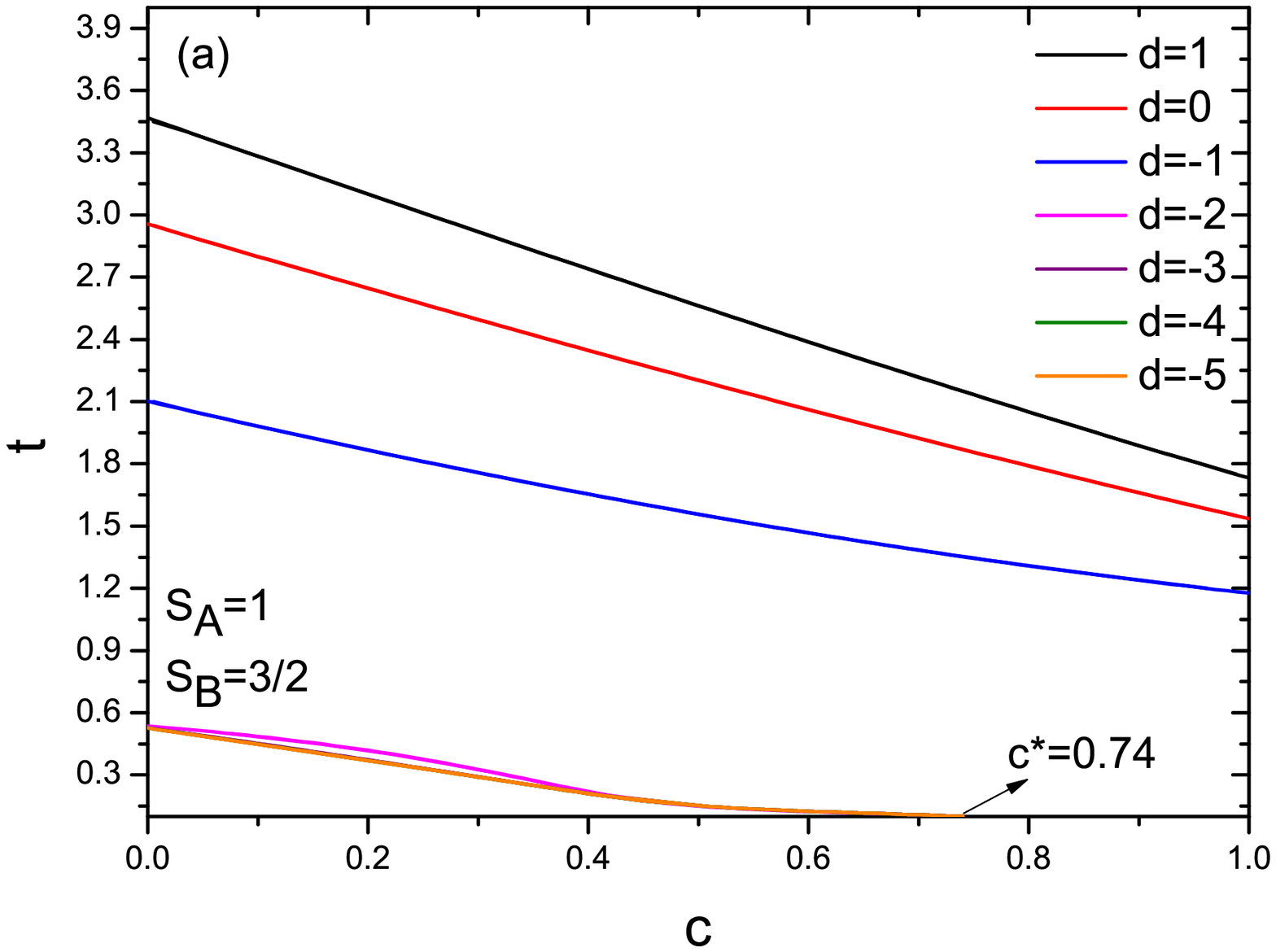, width=8cm}
\epsfig{file=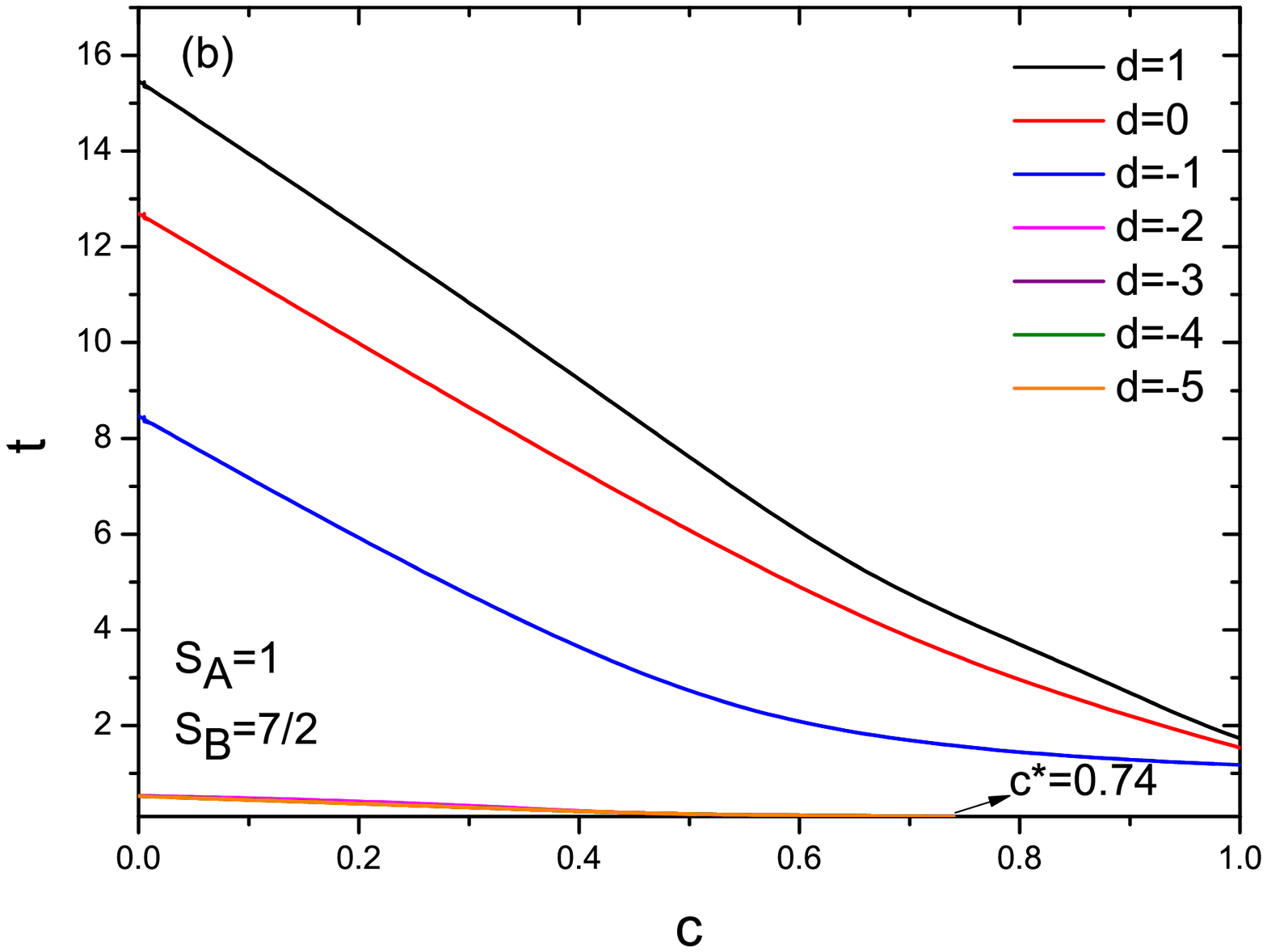, width=8cm}

\centering
\epsfig{file=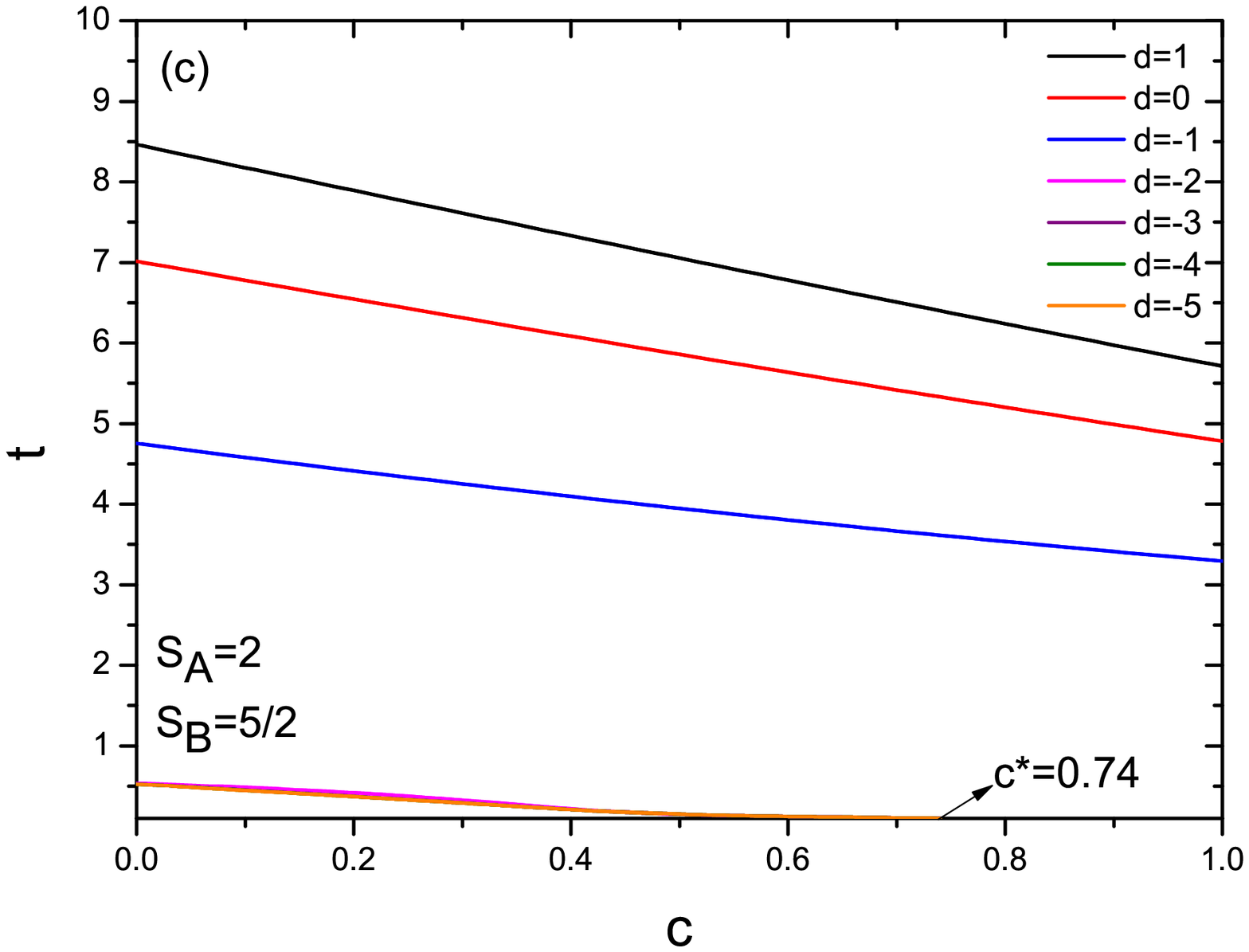, width=8cm}
\caption{Variation of the critical temperature with the concentration for 
selected 
values of crystal field parameters $d=-1$, $d=-2$, $d=-3$, $d=-4$, $d=-5$. Spin 
values of type-A and type-B atoms chosen as (a) $S_A=1$, $S_B=3/2$, (b) 
$S_A=1$, $S_B=7/2$ and (c) $S_A=2$, $S_B=5/2$.
Solid lines represent to the second order transitions.}\label{sek4}
\end{figure}

There are significantly noteworthy results for integer-half integer spin model. 
If we choose spin value of A atoms integer and spin value of B atoms 
half-integer, we obtain different results from the previous cases. 
We depict the phase diagrams in Fig. \ref{sek3} corresponding to (a) $S_{A}=1$ 
$S_{B}=3/2$, (b) $S_{A}=1$ $S_{B}=7/2$ and (c) $S_{A}=2$ $S_{B}=5/2$ spin 
models. 
As we can see from Fig. \ref{sek3} 
the system exhibits  half integer spin valued phase diagram 
for $c=0$ and integer spin valued phase diagram behavior 
for $c=1$, as expected.  Due to the fact that spin value of B atoms greater than the spin value of A atoms, 
as $c$ decreases, the critical temperatures rises. We can see from Fig \ref{sek3} (a), for the concentration 
values $c=0$ and $c=0.5$, the phase diagrams contain second order transition 
lines. The system exhibits tricritical point TCP for $c=1$ and ferromagnetic 
ordered phase destroyed at negative large values of $d$ at the ground state. We can say that, this model behaves like half integer spin valued phase diagram for $c=0.5$. The 
decline of all critical temperatures is valid for all crystal field values for 
$c=0.5$. Note that, the critical temperatures for $c = 0.5$ and $c = 1$  do not merge 
(like in Fig \ref{sek1}  (a)) when the system takes negative large crystal field values. 
If we fix the spin value of A atoms as $S_{A}=1$ and then increase spin value of B 
atoms such as $S_{B}=7/2$, then the critical temperatures also increases for the 
values lower than $c = 1$, which can be seen in Fig \ref{sek3}  (b). If we increase the 
difference between the spin values of A and B atoms, critical lines gradually 
begin to move further apart from each other. This fact can be seen by comparing phase diagrams in Figs. \ref{sek3} (a) and (b). So can we make a general 
inference for concentration? For which value of the concentration, binary 
alloy system behave like integer spin valued phase diagram? To answer these questions, we examined the 
variation of the critical temperature with the concentration values in detail. The phase diagram in $(c,t)$ plane for selected values 
of crystal field parameters $d=1$, $d=0$, $d=-1$, $d=-2$, $d=-3$, $d=-4$ and 
$d=-5$ can be seen in Fig. \ref{sek4}. These curves are constructed for (a) $S_{A}=1$ $S_{B}=3/2$, (b) $S_{A}=1$ $S_{B}=7/2$ and (c) 
$S_{A}=2$ $S_{B}=5/2$ spin models. Critical temperature 
values differ from each other at the limiting cases $c=0$ and $c=1$.  When $c$ 
concentration value increases from 0 to 1, all critical temperatures decrease. 
The system exhibits ferromagnetic phase at low temperatures for all 
concentrations and exhibits paramagnetic phase at high temperatures for 
$d=1$, $d=0$, $d=-1$. If the crystal field parameter takes negative 
values, all critical temperatures decrease. The system exhibits only second 
order phase transition lines for selected values of $d$. It is important to 
emphasize that when the crystal field takes negative large values, the 
concentration value which is the border between the ordered and disordered phase 
at ground state are the same. This critical concentration value is $c^{*}=0.74$ 
which can be seen in Figs. \ref{sek4} (a), (b) and (c). When 
majority of the spins takes integer
values, ground state changes from $s_B=\pm 1/2$ states to $s_A=0$ state. Therefore, 
phase transition occurs at $c^{*}=0.74$ for $d \rightarrow -\infty$ and it exhibits 
disordered phase for values greater than this value.
If we fix the spin value of A atoms as $S_{A}=1$ and 
then increase spin value of B atoms such as $S_{B}=7/2$, then the critical 
temperatures also increases as $c$ goes towards zero (see Fig \ref{sek4} (b)). No matter 
how big the difference between these spin values is, the critical concentration 
value is again $c^{*}=0.74$. As seen in Fig \ref{sek4} (c), again the critical 
concentration value is $c^{*}=0.74$ for 
$S_{A}=2$ $S_{B}=5/2$ spin model. We have also performed calculations for other possible pairs of integer - half 
integer 
spin models, such that $S_{A}<S_{B}$, and we obtain exactly same result. 
$c^{*}=0.74$ value is obtained for all cases.

\subsection{Half integer - Integer spin model}

\begin{figure}[!h]
\epsfig{file=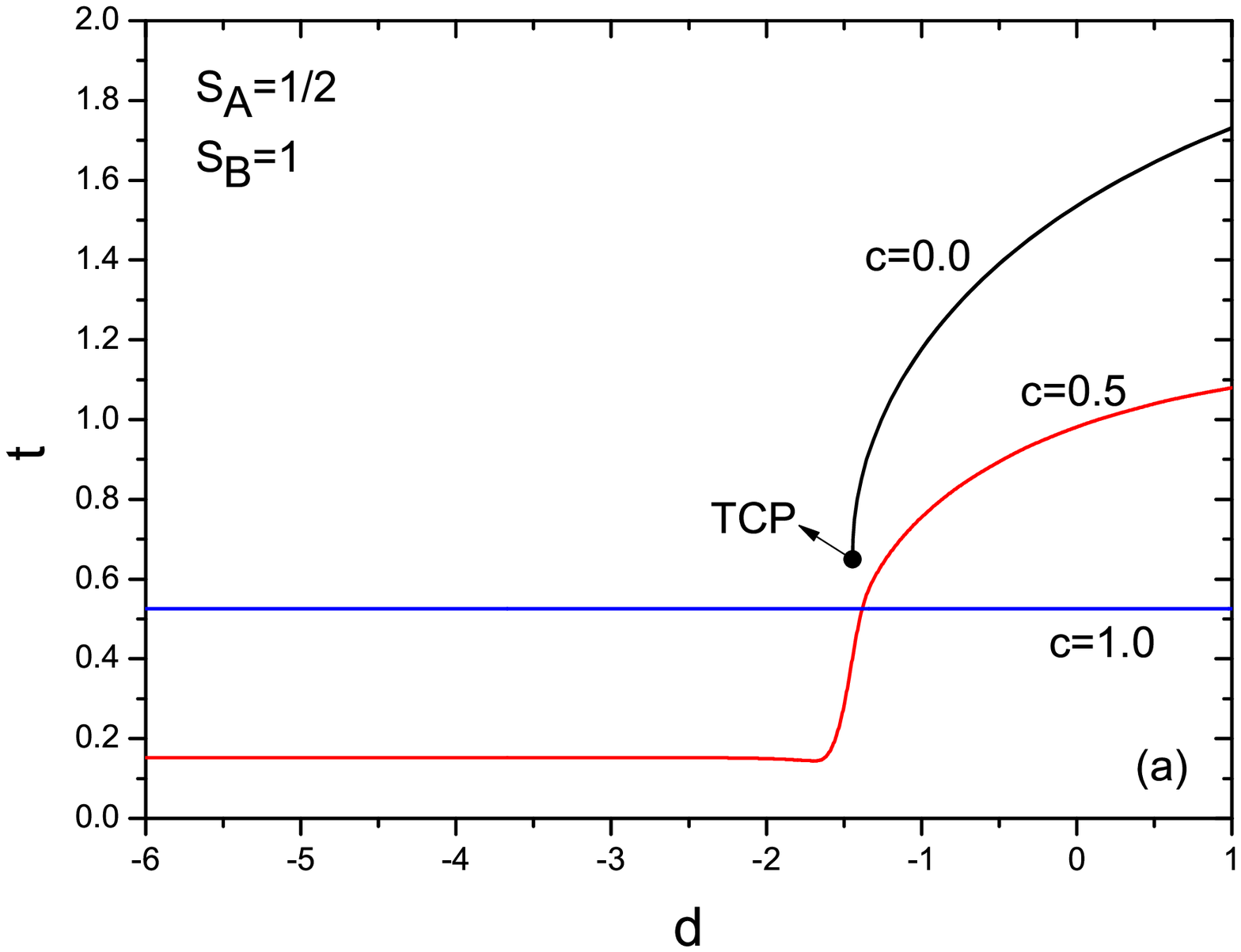, width=8cm}
\epsfig{file=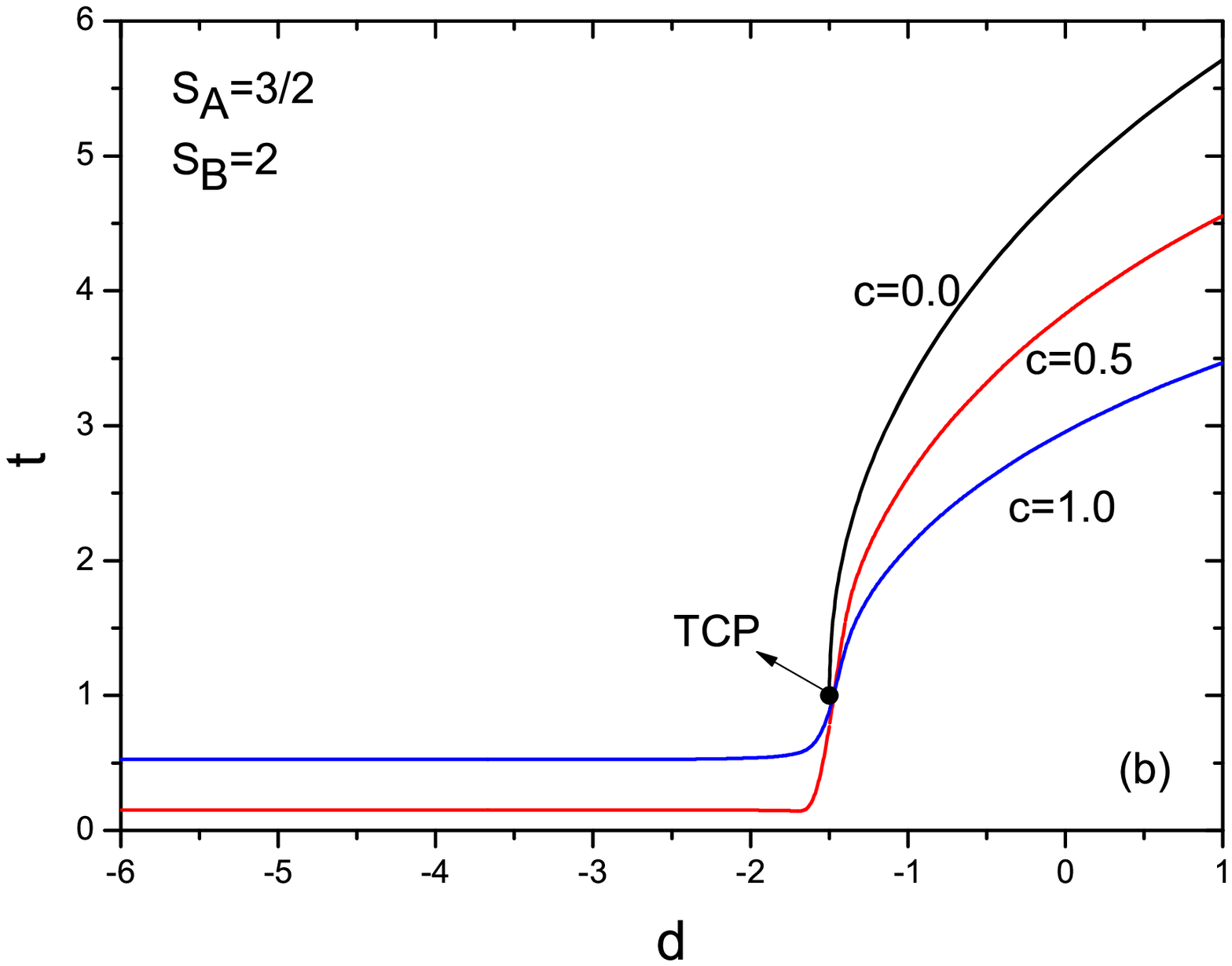, width=8cm}

\centering
\epsfig{file=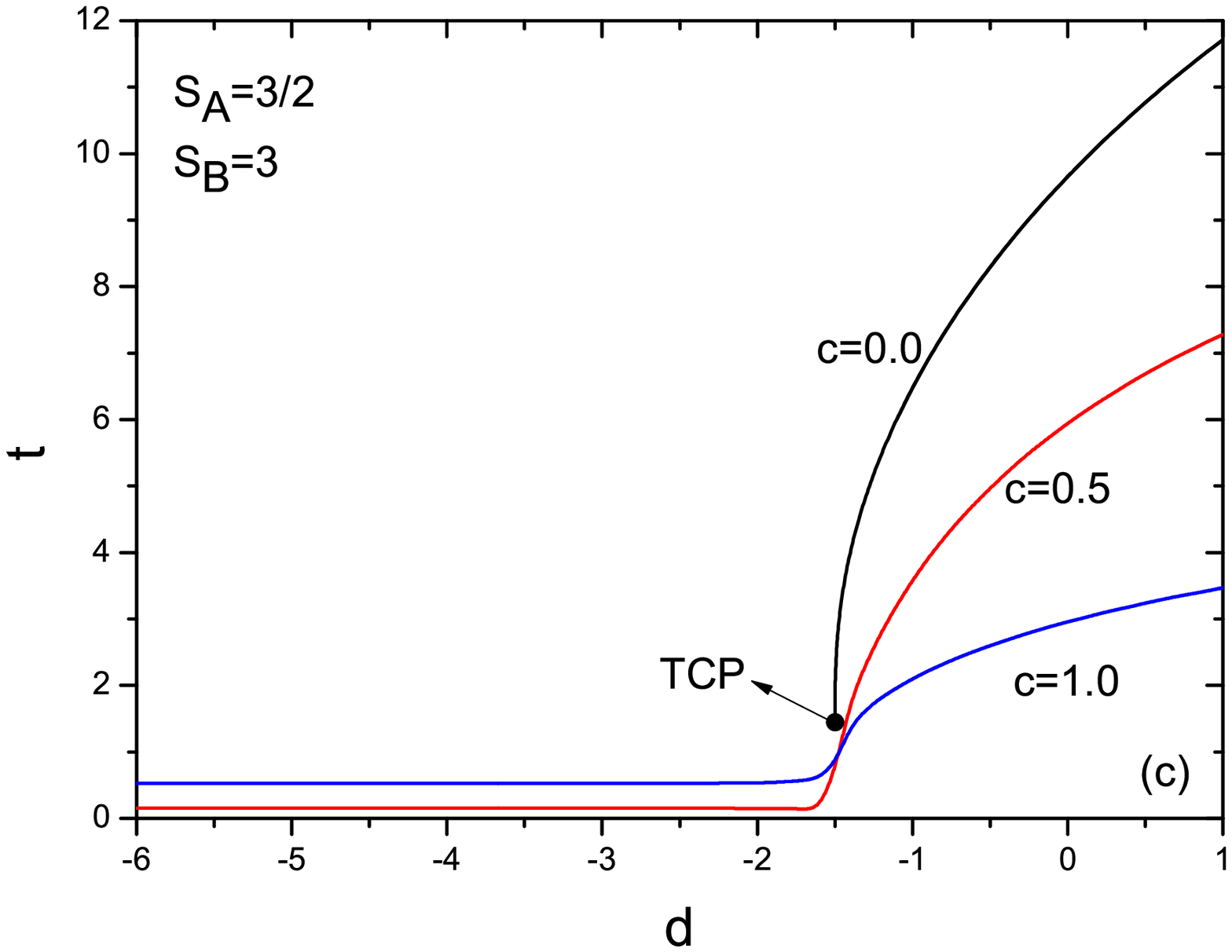, width=8cm}
\caption{Variation of the critical temperature with crystal field parameter for 
selected 
values of the concentration $c=0.0$, $c=0.5$, $c=1.0$. Spin values of type-A 
and type-B atoms are chosen as (a) $S_A=1/2$, $S_B=1$, (b) $S_A=3/2$, $S_B=2$ and 
(c) $S_A=3/2$, $S_B=3$.
Solid lines represent the second order transitions. TCP stands for 
the tricritical point.}\label{sek5}
\end{figure}

\begin{figure}[!h]
\epsfig{file=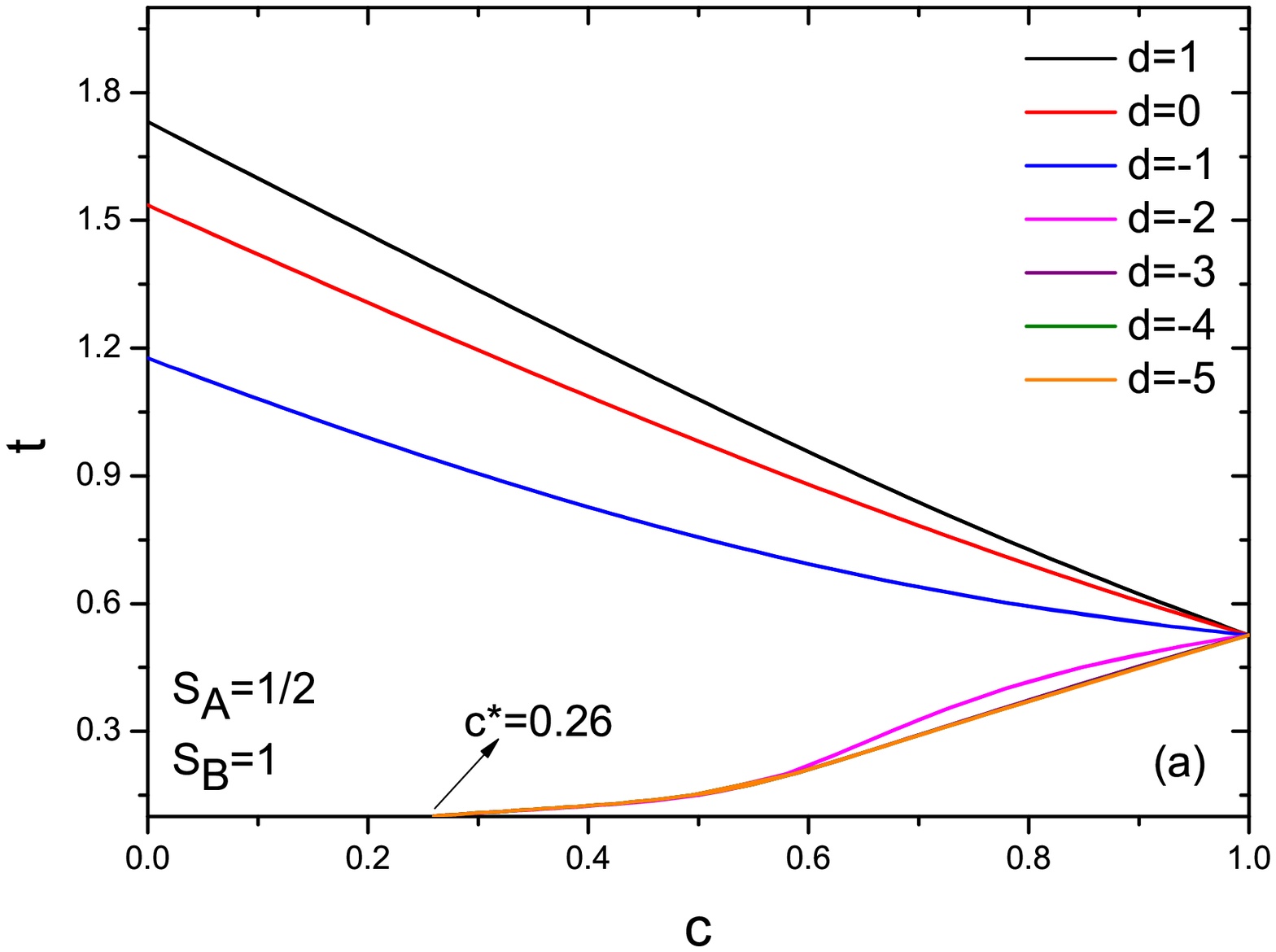, width=8cm}
\epsfig{file=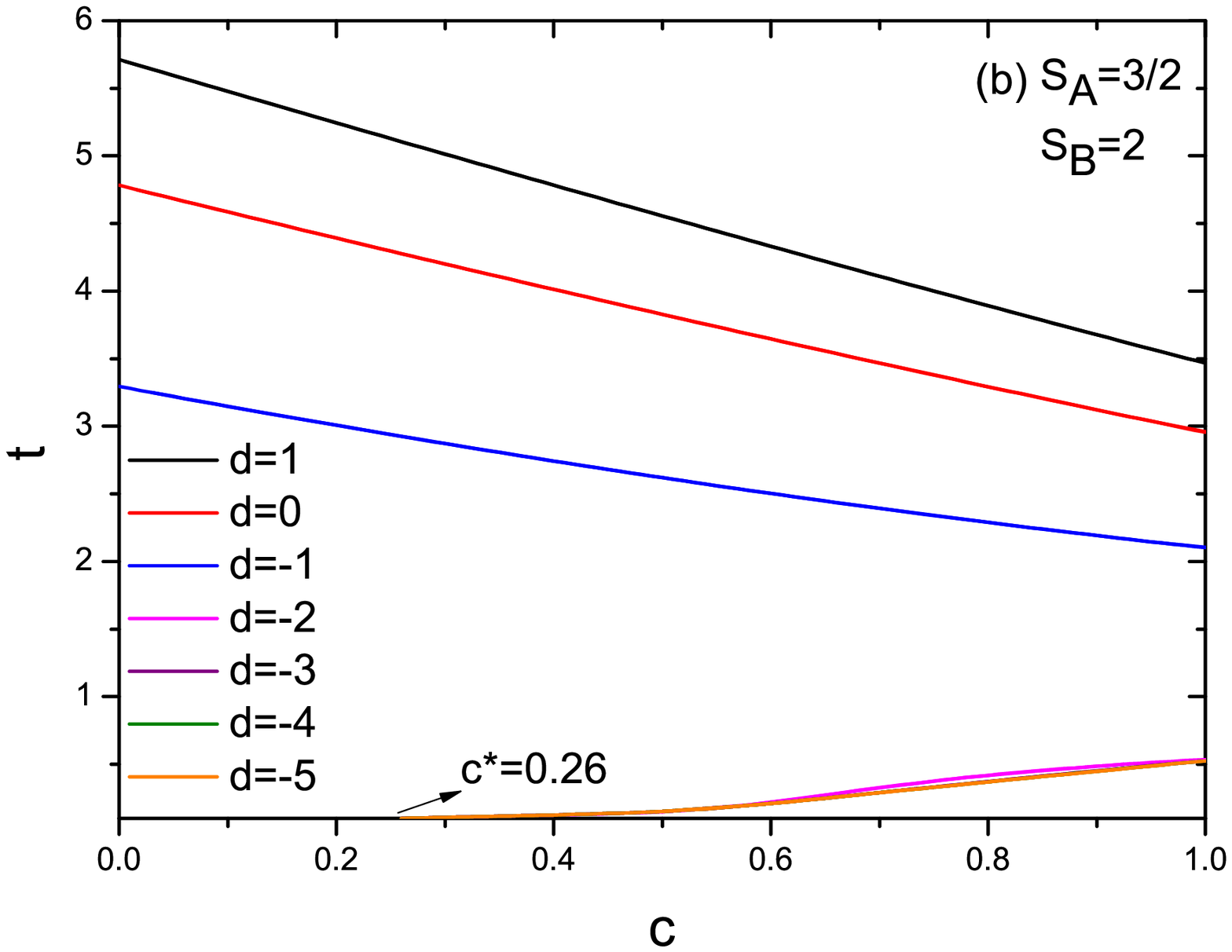, width=8cm}

\centering
\epsfig{file=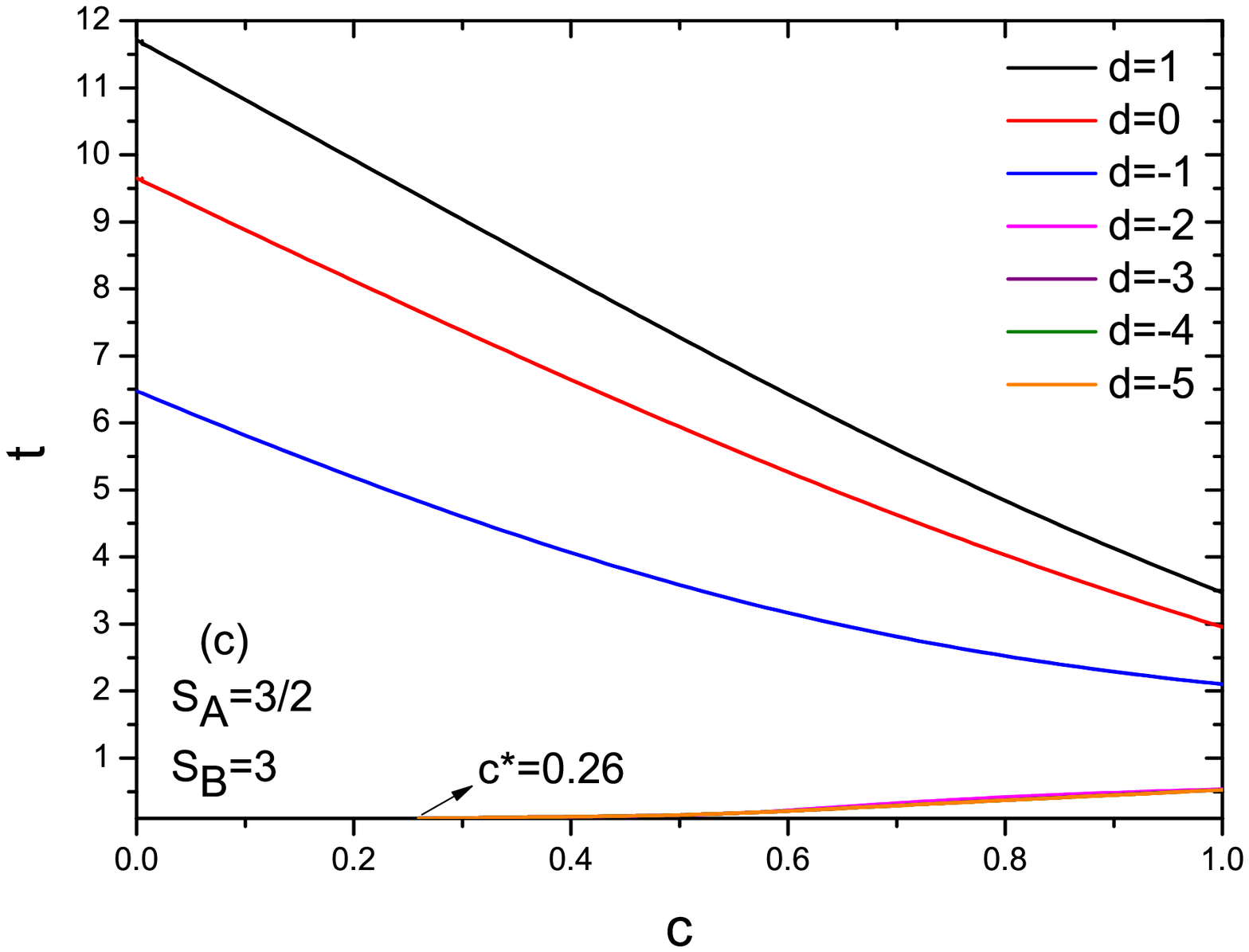, width=8cm}
\caption{Variation of the critical temperature with the concentration for 
selected 
values of crystal field parameters $d=-1$, $d=-2$, $d=-3$, $d=-4$, $d=-5$. Spin 
values of type-A and type-B atoms are chosen as (a) $S_A=1/2$, $S_B=1$, (b) 
$S_A=3/2$, $S_B=2$ and (c) $S_A=3/2$, $S_B=3$.
Solid lines represent the second order transitions.}\label{sek6}
\end{figure}

Although it seems that this subsection is the same as previous subsection, indeed this case is different from the previous one due to $S_A<S_B$.  
We depict the phase 
diagrams of (a) $S_{A}=1/2$ $S_{B}=1$, (b) $S_{A}=3/2$ $S_{B}=2$ and (c) 
$S_{A}=3/2$ $S_{B}=3$ spin models, in Fig.  \ref{sek5}.  As seen in Fig \ref{sek5} (a), the critical temperatures of the system 
will not change along the crystal field plane for $c=1$, due to the fact that, all spins has the value of $1/2$.  The ordered 
ferromagnetic phase area resides below this temperature. As long as the system 
contributes to concentration from the B atoms, ordered phase area expands to the 
higher temperatures for the large values of crystal field parameter. Conversely, 
critical temperature decreases as the crystal field parameter takes negative large 
values. If the majority of concentration consists of integer spin model, ordered 
phase is destroyed under the influence of the negative large crystal field parameters. 
The second order phase transitions turn into the first order transitions, and the system 
displays a TCP. Topology of the phase diagrams is analogous to the integer-half integer binary alloy 
models \cite{ref49,ref50,ref51_2}. If we choose spin values of the system as $S_{A}=3/2$, $S_{B}=2$ 
(Fig \ref{sek5} (b)), the critical temperature increases for larger 
values of crystal field, whereas it decreases for negative large crystal field parameters. 
When all of the lattice sites consist of type-B atoms it displays a larger TCP 
value (compare Fig \ref{sek5} (a) and (b)). In figure \ref{sek5} (c), selected spin values 
are $S_{A}=3/2$, $S_{B}=3$. For higher spin models, if we fix the spin value 
of A atoms as $S_{A}=3/2$ and then increase spin value of B atoms such as 
$S_{B}=3$, both the critical temperatures and TCP increases for positive values of 
$d$. We will ask the first question that comes to our mind. Does the system have 
critical concentration value (as in the previous subsection) in the half integer-integer spin models? 
In order 
to generalize the results about the system, we investigate the behavior of the critical point in  $(c,t)$ plane, for different crystal 
field parameter values.   The phase diagram in $(c,t)$ plane for 
selected values of crystal field parameters $d=1$, $d=0$, $d=-1$, $d=-2$, 
$d=-3$, $d=-4$ and $d=-5$ can be seen in Fig. \ref{sek6}. The variation of the temperature 
with the concentration is constructed for (a) $S_{A}=1/2$, $S_{B}=1$, (b) $S_{A}=3/2$, 
$S_{B}=2$ and (c) $S_{A}=3/2$, $S_{B}=3$ half integer - integer spin models. In 
Fig. \ref{sek6} (a) critical temperatures do not change according to the crystal field 
parameters for the system which has $c = 1$ concentration. This is due to the spin value of $1/2$, which is not affected by crystal field. However, the type-B atoms are 
added to the system, the phase transition lines exhibit higher 
critical temperatures for large crystal field values, while the critical 
temperatures begin to decrease for negative large crystal field values. In other words, the 
ordered phase region increases for large crystal field values, while disordered 
phase appear for negative large crystal field values for the low concentration 
values. The critical concentration value is $c^{*}=0.26$ which yields border 
value between the ordered and disordered phase at ground state. When we start adding 
A atoms to the system, ground state changes from $s_B=0$ states to $s_A=\pm 1/2$ state. Therefore, 
phase transition occurs at $c^{*}=0.26$ for $d \rightarrow -\infty$ and it starts to exhibit 
ordered phase. Thus, we can say 
that the behavior of the integer spin-valued phase diagram is transformed 
into the half-integer spin-valued phase diagram after this critical value. 
In Fig. \ref{sek6} (b) we select $S_{A}=3/2$, $S_{B}=2$ spin model. The system exhibits second order 
phase transitions for all crystal field parameters. Critical temperatures take 
different values at large values of crystal field parameter for $c=1$. 
Ferromagnetic phase area increases as the concentration value goes towards from $1$ to 
$0$ at positive crystal field parameters. The binary alloy model has paramagnetic 
phase area for the low concentrations, in the  presence of the negative large crystal field. 
Critical concentration value appears at $c^{*}=0.26$. For higher order spin 
models, we select $S_{A}=3/2$, $S_{B}=3$ spin model that can be seen in Fig. \ref{sek6} (c). As 
we increase the spin of the B atom, negative large crystal field causes the 
system drags to dragged into the  disordered phase for low concentration values. Similarly, 
$c^{*}=0.26$ is also observed in Fig \ref{sek6}  (c). We have also performed further calculations for all half 
integer- integer spin models for cases of $S_{A}<S_{B}$ which will not be further included  
here but inspected. Therefore, the 
critical concentration value $c^{*}=0.26$ is valid for all half integer- integer 
spin models.

\section{Conclusion}\label{conclusion}

In conclusion, we have investigated the critical properties of generalized spin-S 
magnetic binary alloys represented by $A_{c}B_{1-c}$ within the framework of 
EFT. The system consists of type-A and type-B atoms with the concentration 
$c$ and $1-c$ respectively. The evolution of the phase diagrams in $(d,t)$ and 
$(c,t)$ planes have been presented for selected values of the spins. We have 
obtained that different binary alloy models 
consisting of half integer or integer spin models has its specific phase 
diagram characteristic, as comparable with the literature \cite{ref_9_1,ref21,ref27,ref47,ref49}.

For the binary alloy system which consist of only half integer spins, the system has an ordered 
phase in the ground state, regardless of the value of the crystal field. As expected, the ground state for large negative crystal field region, consists of  occupied $\pm 1/2$ states. All phase diagrams of these systems consist of only second order critical points.   

For the binary alloy system which consist of only integer spins, the system 
has disordered phase for large negative values of the crystal field. This disordered phase has ground state  with filled $0$ states by the spins. Phase diagrams in $(d,t)$ plane exhibit tricritical point TCP. The presence of the TCP is the sign of the first order transitions. All of the phase diagrams 
have TCP regardless of the spin and concentration values, for this case. 

Results of generalized spin-S binary alloy model has also been discussed as half 
integer-integer and integer- half integer spin models. We restrict our results 
in the case $S_{A}<S_{B}$. If the majority of the system consists of 
half-integer (integer) spin values, system exhibit ordered (disordered) phase 
at negative large crystal field values. The phase diagrams may display TCP according to the value of the concentration.  

Another noteworthy point of our results is the presence of exactly the same critical concentration value that drives  the system from  paramagnetic to ferromagnetic phase, regardless of the spin values at large negative crystal field values.  This value is $c^{*}=0.74$ for integer-half integer model and  $c^{*}=0.26$ for half integer - integer model. Note that $(S_A<S_B)$ is chosen. In other words, the integer-half integer binary system cannot have ordered phase for $c>0.74$, for large negative values of the crystal field. In the  same manner   the half integer-integer binary system cannot have ordered phase for $c>0.26$, for large negative values of the crystal field. 

The physical explanation is as follows: 
when the majority of the lattice consists of integer spins,
the ground state should be consists of occupied  $0$  states. 
When the concentration rises in the integer-half integer case, this means that the number of integer 
spins rises. Ferromagnetic interactions between the spins tend the neighbor spins align along the same direction. This  could create ordered phase.  But, after some specific value of this concentration, ferromagnetic interactions cannot create ordered phase due to the rising concentration of occupied $0$ states.

We hope that the results  obtained in this work may be beneficial form both 
theoretical and experimental point of view.





\bibliographystyle{model1-num-names}
\bibliography{<your-bib-database>}







\end{document}